\documentclass[sigconf,nonacm]{acmart}

\usepackage{xeCJK}
\usepackage{booktabs}
\usepackage{tabularx}
\usepackage{longtable}
\usepackage{booktabs}

\renewcommand\footnotetextcopyrightpermission[1]{}

\AtBeginDocument{%
  \providecommand\BibTeX{{%
    \normalfont B\kern-0.5em{\scshape i\kern-0.25em b}\kern-0.8em\TeX}}}

\setcopyright{acmlicensed}
\copyrightyear{2026}
\acmYear{2026}
\acmDOI{XXXXXXX.XXXXXXX}

\usepackage{booktabs} 
\usepackage{url}

\usepackage{caption}
\usepackage[all]{nowidow}
\usepackage{wrapfig}
\usepackage{array}
\usepackage{arydshln}
\usepackage{tabularx}
\usepackage{multirow}
\usepackage{arydshln}
\newcolumntype{L}[1]{>{\raggedright\let\newline\\\arraybackslash\hspace{0pt}}m{#1}}
\newcolumntype{C}[1]{>{\centering\let\newline\\\arraybackslash\hspace{0pt}}m{#1}}
\newcolumntype{R}[1]{>{\raggedleft\let\newline\\\arraybackslash\hspace{0pt}}m{#1}}

\usepackage{wrapfig,lipsum,booktabs} 

\def\authnotes{1}
\newcounter{notectr}[section]
\newcommand{\thenote}{\thesubsection.\arabic{notectr}\refstepcounter{notectr}}

\newcommand{\note}[2]{$\ll$#1~\thenote: #2$\gg$}
\newcommand{\cnote}[1]{\ifnum\authnotes=1 \textcolor{blue}{\note{Comment:}{#1}}\fi}

\begin{document}




\title[Mask, Persona, \& Livestreaming]{Assembling, Breaking, and Refusing the Mask: Agency in AI-Mediated Self-Presentation in Livestreaming}

\author{Yang Hong}
\affiliation{%
  \institution{University of Illinois Urbana-Champaign}
  \city{Champaign}
  \state{Illinois}
  \country{United States}}
\email{yangh9@illinois.edu}

\author{Nusrat Jahan Mim}
\affiliation{%
  \institution{University of Toronto}
  \city{Toronto}
  \country{Canada}}
\email{nusrat.mim@daniels.utoronto.ca}

\author{Sharifa Sultana}
\affiliation{%
  \institution{ University of Illinois Urbana-Champaign}
  \city{Champaign}
  \state{Illinois}
  \country{United States}}
\email{sharifas@illinois.edu}

\renewcommand{\shortauthors}{Hong et al.}

\begin{abstract}
Our mixed-method study examines how Chinese women livestreamers use \textit{masking} to construct idealized mediated personas while navigating gendered, commercial, organizational, and platform pressures alongside personal agendas. We built on the concept of masking, analyzed 627 recruitment posts, and conducted livestream observations and interviews with 26 Chinese women streamers. We found that streamers assembled masks across bodies, AI-mediated technologies, spaces, performances, and social relations to become recognizable while protecting personal boundaries. These masks were continually negotiated, and participants sometimes broke, resisted, or refused them when demands became misaligned or unsustainable. We conceptualize masking as a sociotechnical assemblage in which agency lies in preserving, disrupting, and reconfiguring relations rather than controlling a single interface. We further theorize breaking as a consequential part of masking that exposes hidden labor and unequal costs of visibility. We offer theoretical and design directions for more negotiable, contestable, and agency-supporting AI-mediated self-presentation.


\end{abstract}

\ccsdesc[500]{Human-centered computing~Empirical studies in collaborative and social computing}
\ccsdesc[500]{Human-centered computing~Social media}

\keywords{Livestreaming, masking, self-presentation, gendered visibility, AI-mediated self-presentation, women streamers}


\settopmatter{printfolios=true}

\maketitle

\section{Introduction}


\textit{Masking} is a crucial component in constructing Chinese women livestreamers' recognizable on-camera personas (\textit{renshe},人设) and in navigating visibility. The choice of their online persona requires being supportive of their intended agenda. For example, streamers in brand-promoting businesses are expected to excel in appearance, style, and on-camera performance, following mainstream feminine aesthetics such as sweet, slim, and fair, as per explicit evaluation criteria from industry recruiters and companies. On top of that, streamers seek to hold their own agency on how much of themselves they want to reveal, how they relate to audiences, and what boundaries they maintain around privacy, safety, and identity. Masking practices emerge where these industry expectations and personal agendas intersect. Streamers assemble their masks through the use of beauty filters, AI avatars \cite{wan2024vtubing, liang2026cloning, lu2021kawaii}, as well as styling, props, physical surroundings, and strategic performance \cite{Mim2022fcommerce, XIN2023103360, wang2024harnessing, Wu2023strategy} to become more visible and recognizable on camera.

Prior research has examined several dimensions of technologically mediated presentation in livestreaming, including how beauty filters transform appearance and identities \cite{he2024face,gulati2024beautiful,you2024competitive, freeman2020streaming}, how AI avatars and virtual personas reconstruct authenticity and relationship with audience \cite{wan2024vtubing, lu2021kawaii, Hu2026Beyond, Yin2025vtuber}, and how platform algorithms structure streamers' labor and visibility \cite{xiao2026algorithmic, Bao2025pk, liu2026emotional}. For women streamers, greater bodily visibility brings more attention and monetization opportunities, while it can also expose women to sexualization, harassment, and extra emotional labor \cite{ruberg2019titty,sun2026gender,ancionesanguita2024sexualized, Sanne2025labor}. It also increases exposure of personal information and offline relationships, raising streamers' concerns about managing privacy boundaries \cite{Wu2022concerned,tran2024room}. The sophisticated structure in the Chinese livestreaming industry further complicates these risks \cite{zhang2019nuzhubo, zhang2019livestreaming, Ye2023Navigating}. In this paper, we use \textit{industry} to refer collectively to brands, factories, e-commerce companies, Multi-Channel Networks (MCNs), livestreaming guilds (\textit{gonghui}), agencies, and other intermediaries that recruit, contract, train, or manage streamers \cite{liu2023zhibo,xiao2026algorithmic,yu2025satellite}. Livestream platform algorithms and moderation constrain which personas can remain visible \cite{liu2025violations, Qiao2026live, zhang2026tiered}, while the industry translates commercial and traffic-conversion goals into requirements for how women should appear and perform \cite{xiao2026algorithmic, ye2023ambivalent}.

Treating these technical, commercial, and social pressures as separate problems can lead livestreaming systems to optimize women's visibility without accounting for how they jointly shape the personas women can safely and sustainably present.
We use the concept of \textit{masking} to uncover these relational practices and identify what remains unsupported in the available technologies. Masking derives from Chinese ritual and theatrical performances, where masks and mask-like facial designs construct identities through shared visual conventions. In \textit{Nuo} rituals, masks work with bodily movement and costume to embody the power and expertise needed to conduct ritual ceremonies \cite{li2024nuo,yang2025nuomask}. In Chinese opera, facial makeup (\textit{lianpu}) communicates character traits and social roles through established patterns and colors \cite{chen2016beijingopera,tu2016operamask}. The mask gains meaning through the performer, persona, and audience. We extend this idea to livestreaming, where women use AI beauty filters, styling, space, and performance to build desirable personas while negotiating industry, platform, and audience expectations.

Our mixed-methods study analyzed 627 female livestreamer recruitment posts from chat groups to identify recurring industry requirements and investigated masking practices with 26 Chinese women livestreamers across e-commerce and entertainment through observations and semi-structured interviews. Our research questions were:
\begin{itemize}
   \item[\textbf{RQ1}] How do Chinese women livestreamers assemble and negotiate masks across bodies, AI-mediated technologies, spaces, performances, and organizational and audience relations? How do they follow, negotiate, and reject industry requirements in this process?

    \item[\textbf{RQ2}] How do women streamers break, resist, or refuse their masks within organizational, platform, and audience relations, and how do these negotiations shape their agency over visibility?
    
   \item[\textbf{RQ3}] How can masking practices inform the design of more negotiable AI-mediated self-presentation and extend HCI theories of agency, visibility, and sociotechnical identity?
\end{itemize}
We found that women streamers assembled masks by coordinating bodies, platforms' AI tools, spaces, performances, and social relations around industry-defined ideals of beauty, femininity, and commercial value. They modified company-defined appearances, managed platform and audience expectations, protected personal boundaries, and at times broke, resisted, or refused masking when its demands became untenable. We noted that masking is more than a mere cosmetic or technical functionality, but a sociotechnical site where visibility, agency, and power are negotiated. 

We make four contributions to HCI, Human-AI Interaction (HAI), and Social Media research. \textbf{First}, we provide an empirical and theoretical account of masking as a sociotechnical practice of mediated self-presentation, showing that women streamers assemble masks across bodies, technologies, spaces, performances, and social relations to become recognizable while negotiating boundaries, power, resistance, and refusal. \textbf{Second}, 
we extend assemblage theory to masking by showing that a mediated persona is produced through shifting relations among bodies, technologies, spaces, organizations, audiences, and platforms, reframing agency as the capacity to preserve, disrupt, and reconfigure those relations rather than simply control an AI filter or interface. \textbf{Third,} we theorize breaking as a consequential part of masking rather than its failure. Breakdowns, resistance, and refusal reveal the hidden labor of sustaining a mask and the unequal costs of deviating from or rejecting expected forms of visibility. \textbf{Fourth}, engaging with theories of mask, we offer new theoretical and design perspectives on negotiable AI-mediated self-presentation, agency, contestability, and visibility.
\section{Related Work}
Our study is situated across three bodies of work on mediated self-presentation, gendered visibility, and theories of masking. Together, they motivate our questions about how women streamers assemble, negotiate, resist, and refuse technologically mediated personas, and what these practices contribute to HCI theory and design.

\subsection{Mediated Self-Presentation in Livestreaming}
Livestreaming appears immediate because streamers and viewers connect in real time \cite{maitra2015disaggregating,lu2018youwatch,hu2021slss,hamilton2014streaming}, yet what viewers encounter is already managed and technologically mediated \cite{xiang2025livestreaming,freeman2020streaming}. Goffman's dramaturgical theory frames self-presentation as impression management before an audience \cite{goffman1959presentation}. HCI likewise shows that streamers manage the camera frame, appearance, and interaction, making the visible self socially and technologically constructed \cite{gao2025algorithm,Wu2022concerned,Wang2019greedy}. This self emerges through bodies, technologies, audiences, platforms, and spatial arrangements \cite{Wu2022concerned,Mim2022fcommerce,zou2018producing}, and mediation can be integral to becoming visible and recognizable \cite{wu2025neuro,you2024competitive,Yin2025vtuber}. Research traces increasingly extensive transformations of this visible self. Beauty filters alter facial features and proportions \cite{he2024face} and can reproduce standardized and racialized aesthetic norms \cite{riccio2024mirror,riccio2022openfilter,LIAN2023107793}. Avatars create greater distance from the physical body while retaining awareness of the performer \cite{wan2024vtubing,lu2021kawaii,Yin2025vtuber}; synthetic personas can circulate without the person's immediate presence \cite{xiong2026real}. Face-swapping can detach appearance from participation \cite{deseta2026celebrity}, while Douyin AI avatars reproduce creators' voices, personas, and interaction styles in their absence \cite{liang2026cloning}.

Research on authenticity similarly rejects a simple divide between an unmediated ``real'' self and technological transformation. Audiences can experience AI and virtual streamers as coherent through sustained performance and interaction \cite{hu2026chatbot,wu2025neuro}, while Chinese platforms regulate authenticity as a condition of legitimate creativity \cite{liang2026cloning,Wu2023strategy}. Mediated identities can therefore remain meaningful and even enable more candid expression \cite{wan2024vtubing,lu2021kawaii}. However, this literature mainly explains how technologies transform and represent identities. It says less about how streamers assemble a recognizable self across technologies, bodies, spaces, performances, and social relations, or how they negotiate the expectations attached to that self. These questions are especially consequential for women, whose mediated appearances are shaped by gendered expectations of visibility, attractiveness, authenticity, and appropriate self-presentation \cite{freeman2020streaming,sun2026gender,ruberg2019titty,zhang2019nuzhubo}. This motivates \textbf{RQ1's} focus on how masks are assembled and negotiated, while providing the basis for \textbf{RQ3's} theoretical and design inquiry into more negotiable AI-mediated self-presentation.

\subsection{Gendered Visibility and Livestreaming}

Visibility in livestreaming is not socially neutral, particularly for women whose bodily appearance can simultaneously attract audiences and invite gendered scrutiny \cite{ruberg2019titty,sun2026gender,Zhen2024pretty}. Women streamers may gain visibility while facing harassment, sexualization, and pressures to perform recognizable forms of femininity \cite{freeman2020streaming,sun2026gender,tang2022dare,nakandala2017gendered}. Ruberg et al. show how the stigmatizing category of the \textit{titty streamer} can reduce women game streamers to bodily appearance and presumed sexuality \cite{ruberg2019titty}. These negotiations are also culturally situated. Chinese \textit{nüzhubo} are interpreted through gendered and sexualized expectations \cite{zhang2019nuzhubo,zhang2019livestreaming,zimmer2019drives}, while rural women engaged in e-commerce livestreaming negotiate family and community judgments about respectable femininity \cite{tang2022dare}. Visibility can therefore create economic opportunity while also producing stigma, sexualization, and conflicting expectations about how women should appear and behave.

Platform economies further shape which forms of gendered visibility become valuable and permissible. MCNs translate opaque algorithms into training, metrics, and account management that encourage appearances and performances expected to attract traffic \cite{xiao2026algorithmic,liu2023zhibo}. Platforms reinforce these expectations through algorithmic pedagogy \cite{liang2025pedagogy,Qiao2026live,riccio2024exposed} and moderation systems that regulate visible language, imagery, and behavior \cite{liu2025violations,Qiao2026live}. Women must therefore become socially recognizable, commercially valuable, algorithmically legible, and institutionally permissible across overlapping systems of evaluation \cite{Mim2020others,xiang2025livestreaming,wang2024harnessing}. Existing research examines these pressures through self-presentation, privacy, harassment, platform labor, and algorithmic governance \cite{Wu2022concerned,ye2023ambivalent,tran2024room,Ahmed2019privacy,Mim2022fcommerce,Saha2024sex}. What remains less clear is how women respond when the forms of visibility demanded by these systems conflict with their own boundaries or become difficult to sustain. This motivates \textbf{RQ1's} attention to how women negotiate industry-defined masks and \textbf{RQ2's} focus on when they break, resist, or refuse them, and what these moments reveal about agency over visibility.

\subsection{Theories of Masking: Identity, Recognition, and Power}

Next we turn to the theories and traditions of masking that complicate the common understanding of a mask as something that simply conceals an underlying or authentic self. In Chinese ritual and performance traditions, masks instead participate in producing identities through culturally recognizable signs. In \textit{Nuo} rituals, for example, distinctive masks operate together with bodily movement, costume, and performance to enable performers to embody particular gods, spirits, and ritual figures \cite{li2024nuo, shen2022bodily, yang2025nuomask}. Chinese opera similarly uses mask-like facial makeup (\textit{lianpu}) as a system of visual signification, where colors, patterns, and facial designs communicate aspects of a character's personality, morality, social position, and dramatic role; their intelligibility depends on cultural conventions shared between performer and audience \cite{tu2017cultural}. The mask therefore does more than alter the wearer's appearance: it makes an identity recognizable by placing the wearer within an existing system of signs. Yet the terms of recognition are not neutral. Fanon's analysis in \textit{Black Skin, White Masks}, situated within colonial racial hierarchies, shows how becoming intelligible to others can require subjects to negotiate categories of language, culture, and identity whose meanings and values have already been constituted through unequal social relations \cite{fanon1952black}. Fanon thus shifts attention from what an individual intends to present toward the conditions under which that presentation can be recognized. These perspectives reveal a central feature of masking: the mask makes the wearer recognizable through signs that are never entirely the wearer's own.

However, becoming recognizable through an existing system of signs does not make the person equivalent to the identity those signs prescribe. Bhabha's account of mimicry is useful because mimicry depends upon resemblance while unsettling complete correspondence: recognizable forms can be repeated and inhabited without eliminating the difference between the norm and the subject who performs through it \cite{bhabha1994location}. The resulting ambivalence lies not simply between constraint and agency, but between recognition and difference --- between becoming legible through an established image and remaining irreducible to that image. Riviere's view of ``womanliness as a masquerade'' further complicates this relationship by questioning whether a stable boundary can be drawn between an underlying femininity and femininity performed through socially recognizable signs \cite{riviere1929womanliness}. Masking therefore cannot be reduced to either hiding an authentic self or adopting an artificial one. It describes a relational process in which people become recognizable by inhabiting available images while negotiating their relationship to what those images signify. We bring this understanding to AI-mediated livestreaming, where masks mediate women's recognition within norms of femininity and visibility without reducing them to the images through which they appear. This framing motivates \textbf{RQ1}'s focus on how masks are assembled and negotiated, \textbf{RQ2}'s attention to breaking, resistance, and refusal, and \textbf{RQ3}'s inquiry into how masking can inform HCI theory and design.
\section{Methods}

Our two-phase mixed-methods study combines analysis of 627 live-streamer recruitment posts, and observations and semi-structured interviews (see Table \ref{tab:participant_summary}). Our university's Institutional Review Board (IRB) approved our research. This research was conducted between June and August 2026. 

\subsection{Phase 1: Recruitment Post Analysis}
\subsubsection{Data Collection}
To understand the explicit requirements of the Chinese livestreaming industry, we joined two highly active public Wechat groups dedicated to livestreamer recruitment, with a combined membership of 998. Recruitment information posted in these groups was also commonly cross-posted to other Chinese recruitment platforms, such as Boss Zhipin \cite{bosszhipin2026}. We collected 1,268 recruitment posts shared in these groups over two weeks in July 2026. We used GPT-5 to assist with an initial screening step that removed duplicate posts and posts outside our inclusion criteria, such as advertisements recruiting only male streamers or offering livestreaming equipment rental. The LLM was used only for corpus screening; all substantive coding of recruitment requirements was conducted manually. To evaluate the screening process, we manually audited a stratified random sample of both posts retained and those excluded by the LLM. The first author, a native Chinese speaker, independently assessed whether each sampled post contained recruitment information relevant to women livestreamers. The manually audited sample contained 54 relevant posts and 46 irrelevant or duplicate posts according to the first author's annotations. We specifically examined excluded posts for euphemistic, abbreviated, or niche recruitment language that the model might have missed. Using the first author's manual annotations as the reference, the model achieved 100\% precision and 96.3\% recall, with a false-negative rate of 3.7\%. We manually reviewed screening disagreements and corrected misclassified posts before constructing the final corpus. After screening, the final corpus contained 627 recruitment posts targeting women livestreamers.

\subsubsection{Data Analysis}
We conducted a conceptual content analysis to identify the recurring requirements \cite{krippendorff2004content, Zimmer2018Content}. We treated each recruitment post as a multi-label unit and coded the presence of requirement concepts expressed through words or phrases. A requirement was counted at most once within a post, regardless of how many times it appeared. The same post could contain multiple requirements and therefore contribute to multiple categories. The first author manually coded the full corpus in Chinese to preserve the contextual meanings of industry-specific terms. We began with five provisional themes identified from the first 20 posts and iteratively added categories as new requirements appeared. This process initially produced 12 themes. We then developed a codebook specifying the definition, inclusion and exclusion criteria, and representative expressions for each theme. Categories were not designed to be mutually exclusive. For instance, a phrase describing a streamer as both ``young feminine'' and ``slim'' could be coded under age and appearance when it communicated both requirements. Through iterative comparison, we merged categories that captured the same underlying recruitment criterion. Height and weight, for example, were grouped under \textit{body size} because both specified desired bodily dimensions. This process resulted in eight broader themes. 

\begin{table*}[t]
\centering
\caption{Summary of Different Attributes of the Interview and Observation Participants.} 
\Description{Summary of the 26 interview and observation participants.  The table reports participants' livestreaming types, education levels,  platforms used, and industry relationships. Fourteen had entertainment livestreaming experience, and 15 had product-selling or consultation experience, with three doing both. Douyin was the most commonly used platform, followed by RedNote and WeChat Live. Twelve participants had  worked with a livestreaming guild or MCN.}
\vspace{-10pt}
\label{tab:participant_summary}
\renewcommand{\arraystretch}{1.2}
\setlength{\tabcolsep}{6pt}

\begin{tabular}{|p{0.21\textwidth}|p{0.21\textwidth}|p{0.215\textwidth}|p{0.235\textwidth}|}
\hline
\multicolumn{4}{|c|}{\textbf{Participants ($n=26$, 19--46 years old)}} \\
\hline
\centering
\textbf{Livestreaming Type}\\[1mm] 
\begin{tabular}{rl}
Entertainment: & 14 \\
Product selling/ & \\
consultation: & 15 \\
Both: & 3
\end{tabular}
&
\centering
\textbf{Education}\\[1mm]
\begin{tabular}{rl}
Bachelor's degree: & 8 \\
Master's degree: & 6 \\
Associate degree: & 5 \\
Undergraduate: & 4 \\
High school: & 2 \\
Graduate: & 1
\end{tabular}
&
\centering
\textbf{Platform}\\[1mm]
\begin{tabular}{rl}
Douyin: & 16 \\
RedNote: & 13 \\
WeChat Live: & 6 \\
Kuaishou: & 3 \\
Taobao: & 2 \\
Bilibili: & 2 \\
Other: & 5
\end{tabular}
&
\centering
\textbf{Industry Relations}\\[1mm]
\begin{tabular}{rl}
Guild/MCN: & 12 \\
Brand partnership: & 4 \\
Independent: & 4 \\
Company/Factory: & 3 \\
MCN ownership: & 2 \\
Factory ownership: & 1
\end{tabular} 
\tabularnewline
\hline 
\end{tabular}
\end{table*}

\subsection{Phase 2: Interviews and Observation}
The second phase combined semi-structured interviews (n=26) to examine participants' desired personas, masking practices, motivations, and perspectives on current and desired livestreaming technologies and livestream observations (n=23, 17 hours) to further understand their surrounding contexts.

\subsubsection{Participant Recruitment}
We recruited 26 participants through social media and purposeful sampling (Tab-1). We defined a ``woman livestreamer'' as a user who had a personal account on a livestreaming platform, had experience hosting livestreams, and self-identified as female \cite{tang2022dare}. Participants varied in age from 19 to 46 and used a wide range of popular livestreaming platforms, including Douyin, Kuaishou, Huya, RedNote, WeChat Live, Taobao, JD Live, and Bilibili, each representative of different types of livestreaming content and communities in China. Sixteen participants livestreamed on multiple platforms. We obtained oral consent from all participants. 

\subsubsection{Semi-structured Interviews}
We conducted online semi-structured interviews lasting around 40--60 minutes. Interview questions covered participants' livestreaming preparations, their desired online personas, the ways they presented, masked, or polished themselves and their surroundings, their motivations for doing so, the conditions that shaped these practices, and their perspectives on current and desired livestreaming technologies and AI tools. All interviews were conducted in Mandarin. With participants' consent, interviews were conducted remotely through voice or video calls and audio-recorded for transcription and analysis. We removed personally identifiable information from the transcripts before analysis. During the interviews, some participants also shared photographs of their livestreaming environments, including room setups, lighting equipment, and products displayed during livestreaming.

\subsubsection{Observation}
We also observed the interview participants' livestreaming sessions to further understand their surrounding contexts. For each observation session, we joined a livestream and observed both the streamer's activities and viewers' comments for about 30--60 minutes. During the observations, we focused on streamers' (1) livestreaming settings and backgrounds, (2) use of filters, (3) makeup and styling, (4) livestream content, such as the products they sold, and (5) interactions with viewers and other people in the livestream, such as assistants and streamers joining through \textit{Lianmai} and \textit{PK} sessions. We wrote down detailed notes and captured screenshots of representative moments to support subsequent analysis. To minimize disruption, we did not directly interact with the streamers during the observation sessions. In total, we conducted 17 hours of observations with 23 participants, as three participants were no longer livestreaming at this phase of the study.

\subsubsection{Data Analysis}
We collected approximately 24 hours of interview recordings, observation notes, screenshots of representative moments, and photographs shared by participants. The recordings were transcribed in Mandarin and translated into English by the first author. Back translation was employed to verify the consistency and accuracy. We conducted thematic analysis of the collected data \cite{corbin2014basics,boyatzis1998transforming}, beginning by reading through the interview transcripts and observation notes closely, allowing codes to develop. Forty-eight initial codes spontaneously developed to capture recurring practices and concerns. Through several iterations, we compared and clustered related codes into higher-level themes. Examples of higher-level themes included ``types of persona'', ``filter use'', ``spatial arrangement''. We revisited the observation fieldnotes, screenshots, and participant-shared photographs throughout the analysis to contextualize and refine the themes. We present our findings from observations and interviews in the section \ref{findings}.

\subsection{Positionality}
Our research team consists of three women. The first author was born and raised in China, is a native Mandarin speaker, and is familiar with mainstream Chinese livestreaming platforms. This cultural and linguistic familiarity helped her access participant communities, understand industry-specific language, and build rapport during interviews. At the same time, her position as an HCI researcher placed her outside participants' everyday work as livestreamers. This insider-outsider position shaped the research relationship. The other two authors are non-Chinese and have extensive experience in critical-HCI research. We remained attentive to how this distance could shape our interpretations during data collection and analysis. Because interviews were conducted in Mandarin and findings were written in English, we also considered how translation could flatten culturally and industry-specific meanings. We retained key Chinese terms where their meanings could not be fully captured in English. We acknowledge that our background shapes both the questions we ask and our interpretation of the findings \cite{harrington_deconstructing_2019}, particularly our attention toward agency, gender, and structural power. This study is a first step in a longer-term inquiry into technology use and design with women livestreamer communities.
\section{Chinese Women Livestreamers and Livestreaming Practices}
\label{context}
China's livestreaming industry grew out of public online chat rooms in the mid-2000s. Over the following two decades, livestreaming expanded from gift-driven \textit{showroom} entertainment into a broader ecosystem that increasingly incorporates e-commerce \cite{xiang2025livestreaming,Lu2019responsbility}. Major platforms have integrated shopping and storefront functions into livestreaming. The line between showroom streamers and e-commerce streamers is becoming increasingly blurred \cite{cnnic2025internet}. 

\subsection{Who Are ``n\"{u}zhubo''?}
\subsubsection{Gendered Stigma Around \textit{N\"{u}zhubo}}
Women have been highly visible in livestreaming since the industry's early development. The term \textit{n\"{u}zhubo}, literally ``woman livestreamer,'' became associated especially with entertainment and showroom livestreaming, where women sing, dance, chat, and cultivate viewer relationships, often through virtual gifts \cite{ye2023ambivalent,Zhen2024pretty}. This visibility has also produced persistent stigma. Prior work shows that women streamers are often characterized as hypersexual, vulgar, or dependent on flirtatious relationships with men, even when they emphasize the skill and labor involved \cite{ye2023ambivalent,Zhen2024pretty}. Participants encountered similar assumptions that women streamers earn ``quick money'' through appearance, are poorly educated or materialistic, or engage in sexually suggestive \textit{cabian} content. These stereotypes shaped disclosure. Even when participants considered their content ordinary and platform-compliant, many found the identity of \textit{n\"{u}zhubo} socially uncomfortable. Ten described livestreaming as part-time, temporary, or informal work rather than a long-term professional identity.

\subsubsection{Different Paths into Livestreaming}
Participants entered livestreaming from diverse positions, including students, office workers, actors, shop owners, entrepreneurs, international students, and service workers. Some livestreamed full-time, while others combined it with existing jobs, businesses, or social-media work. Their entry paths also varied. Some were recruited through social media, while others began livestreaming to extend an existing audience or generate income. Livestreaming also emerged through participants' occupations: actors used it for auditions, marketing employees promoted company products, and shop or factory owners used it to increase sales.

\subsubsection{Livestreaming Roles and Industry Relations}
Participants' work fell into two overlapping forms. Fifteen engaged in product selling or consultation, including tarot reading and other services. Two operated factories, and two early Taobao Live streamers later ran their own MCNs. Fourteen had experience with entertainment live-streaming involving chatting, singing, dancing, PK competitions, and real-time viewer interaction. Two had worked in \textit{tuanbo}, where several streamers perform together while viewers support individuals through virtual gifts. Some participants moved between sales and entertainment formats. Across both forms, 24 had worked with companies, brands, factories, agencies, or other intermediaries, showing how livestreaming personas were rarely produced by streamers alone.

\subsubsection{Unequal Visibility and Precarious Work}
Participants' experiences complicated the image of livestreaming as an easy path to visibility and income. Industry terminology distinguishes \textit{toubu zhubo} with over one million followers, \textit{yaobu zhubo} with over 100,000, and \textit{dibu zhubo} with fewer than 100,000 \cite{Ye2023Navigating}. Most participants fell into the \textit{dibu} category; P26 was an exception with about 260,000 followers. Participants described moving beyond the lower tier as difficult in China's crowded market.

\subsubsection{Livestreaming as a Profession} Livestreaming is demanding and uncertain. A national report found that over 80\% of professional streamers earned under RMB 8,000 per month, while 57.4\% of daily streamers broadcast for more than six hours a day \cite{streamerreport2024}. Participants similarly described low base salaries, commission-based income, contractual restrictions, and late-night schedules. P3 noted that six advertised livestreaming hours could become seven or eight with preparation. Newcomers sometimes worked overnight for 
greater traffic. Participants also distrusted some MCNs and guilds because of added hours, low salaries, and unfair shares of virtual-gift income \cite{yejin2021guilds,streamerwork2025}. Together, these conditions made visibility both desirable and difficult to sustain.

\begin{figure*}[t]
  \centering
  \includegraphics[width=0.82\textwidth]{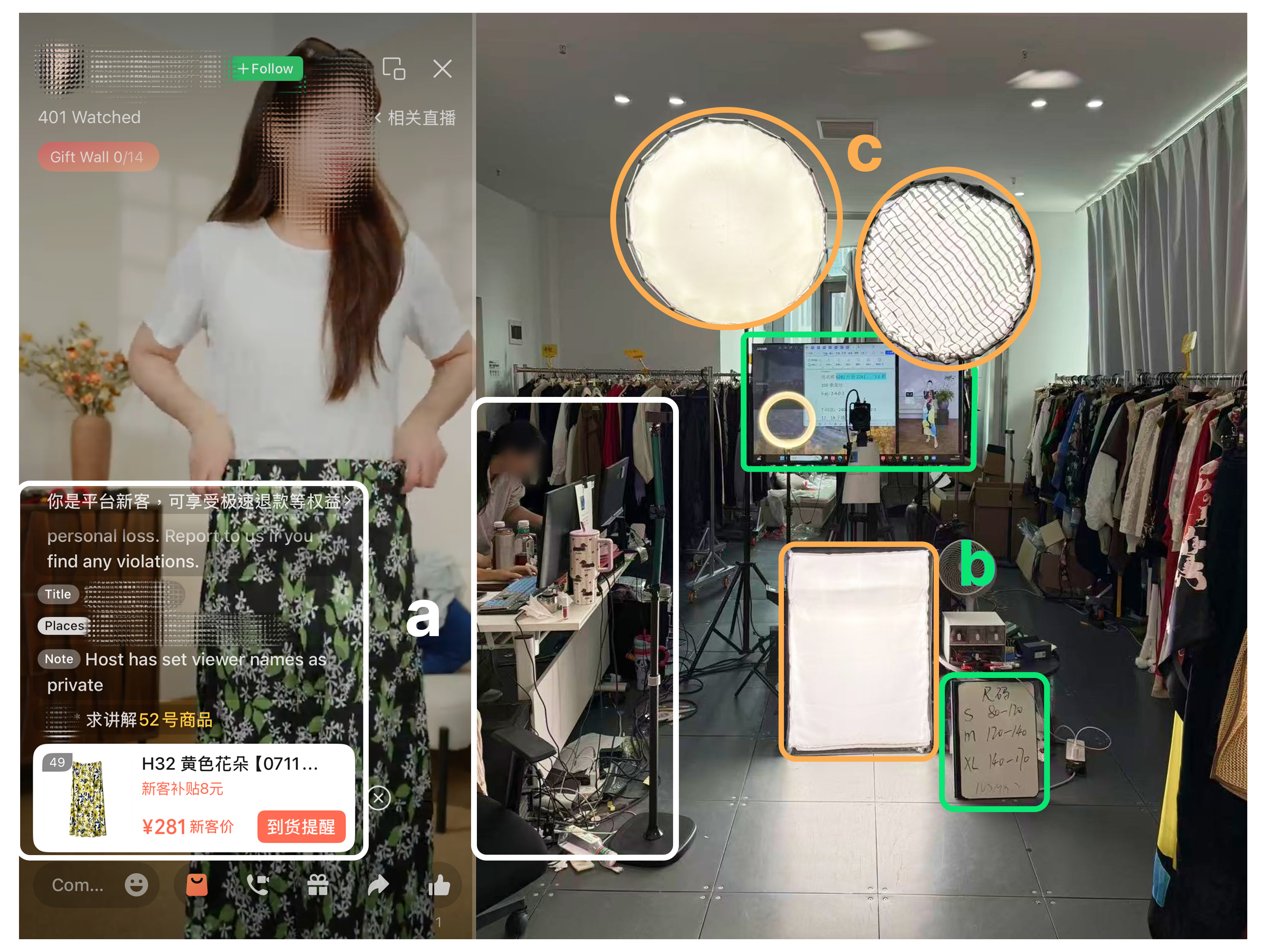} \vspace{-10pt}
  \caption{Front- and backstage views of a commercial livestreaming room. The left image shows the WeChat Live interface, where the streamer presents a dress. The right image shows the production environment outside the audience-facing frame. (a) A central control operator (\textit{zhongkong}) manages product links and monitors viewers' real-time comments (\textit{danmu}). (b) A prompting screen provides cues for the streamer, while a whiteboard below lists sizing information. (c) Multiple lights are positioned around the streamer, with clothing racks and other products surrounding the livestreaming area.}
  \Description{The left image shows a woman streamer presenting a dress through the WeChat Live interface. The right image shows the same livestreaming room from behind the camera, including a central control operator at a desk, clothing racks, multiple lights, a prompting screen, a whiteboard, and other livestreaming equipment.} 
  \label{fig:livestream-room}
\end{figure*}

\subsection{Preparing and Performing a Livestream}
\subsubsection{Preparing the Streamer and the Visible Environment}
The preparation began hours before the camera turned on. Participants prepared makeup, hair, clothing, and styling, while entertainment streamers gathered memes, conversational prompts, and recurring phrases. Product-selling streamers reviewed product information, selling points, ordering procedures, and sales scripts. P20, for example, began two hours before going live to arrange products and verify links, colors, and sizes. Participants also prepared the visible environment by adjusting backgrounds, lighting, camera angles, products, and props. Commercial studios could include clothing racks, product displays, prompting screens, monitors, microphones, mirrors, and multiple specialized lights.

\subsubsection{Coordinating the Livestream Backstage}
Live broadcasting is a coordinated real-time labor. Participants described a formal commercial team with four roles: the streamer, operations staff (\textit{yunying}), a central control operator (\textit{zhongkong}), and an assistant host. The streamer performed on camera, while the assistant host handled comments, questions, products, and transitions. The \textit{zhongkong} managed product links and chat, and operations staff monitored performance and sales data, adjusted pacing, and shaped the account's broader direction. P6 emphasized their influence, noting that operations staff could decide when to drop poorly performing products, how to manage the account, and how the streamer's persona should develop. Four-person teams were uncommon. P6 said three-person teams were more typical, with one person combining roles such as \textit{zhongkong} and operations. Figure~\ref{fig:livestream-room} illustrates this division of labor. Viewers see a woman presenting clothing within the interface, while outside the frame, staff monitor comments and product links, prompting screens provide cues, and lights and merchandise surround the streamer.

\subsubsection{Streaming With and Without Support}
Many participants streamed without this backstage support. Independent streamers had to perform while monitoring comments, managing the platform interface, and handling disruptions themselves. Entertainment streamers sometimes relied on a volunteer \textit{room moderator} (\textit{fangguan}), often a trusted long-term viewer, to manage hostile messages and help maintain the room. Some streamers also worked with a \textit{manager} (\textit{jingjiren}), whose role could extend from recruitment to emotional support. P15 explained, ``Livestreaming is very stressful, and it is easy to overthink things. When I run into problems, my manager may help me work through them.'' Manager and operations roles sometimes overlapped. P12's manager also monitored her livestream from the backend and sent real-time suggestions through highlighted messages.

\subsubsection{Work Beyond the Broadcast}
Livestreaming work continued after the camera turned off. Streamers reviewed metrics such as virtual-gift revenue, sales, viewer counts, viewing duration, audience retention, and demographics, either with operations staff or alone. Entertainment streamers also maintained relationships with high-spending viewers through personal social-media accounts. E-commerce streams could generate additional logistical work. In P5's fabric business, she and her sister alternated between livestreaming and customer service, then printed orders, packed fabric, and prepared shipments after streams that sometimes ended at midnight or 2:00~AM.
\section{Industry-imposed Personas on Women Livestreamers}
\label{findings}

Our analysis of 627 recruitment posts mapped the industry expectations on what a livestreamer should look like. Appearance was a frequently specified theme ($N=641$), followed by on-camera capabilities ($N=561$), sales and conversion capabilities ($N=494$), aesthetic styles ($N=451$), body size ($N=448$), personality ($N=250$), and perceived age ($N=205$) (see Fig.2). 

\begin{figure}[t]
  \centering
  \includegraphics[width=0.5\textwidth]{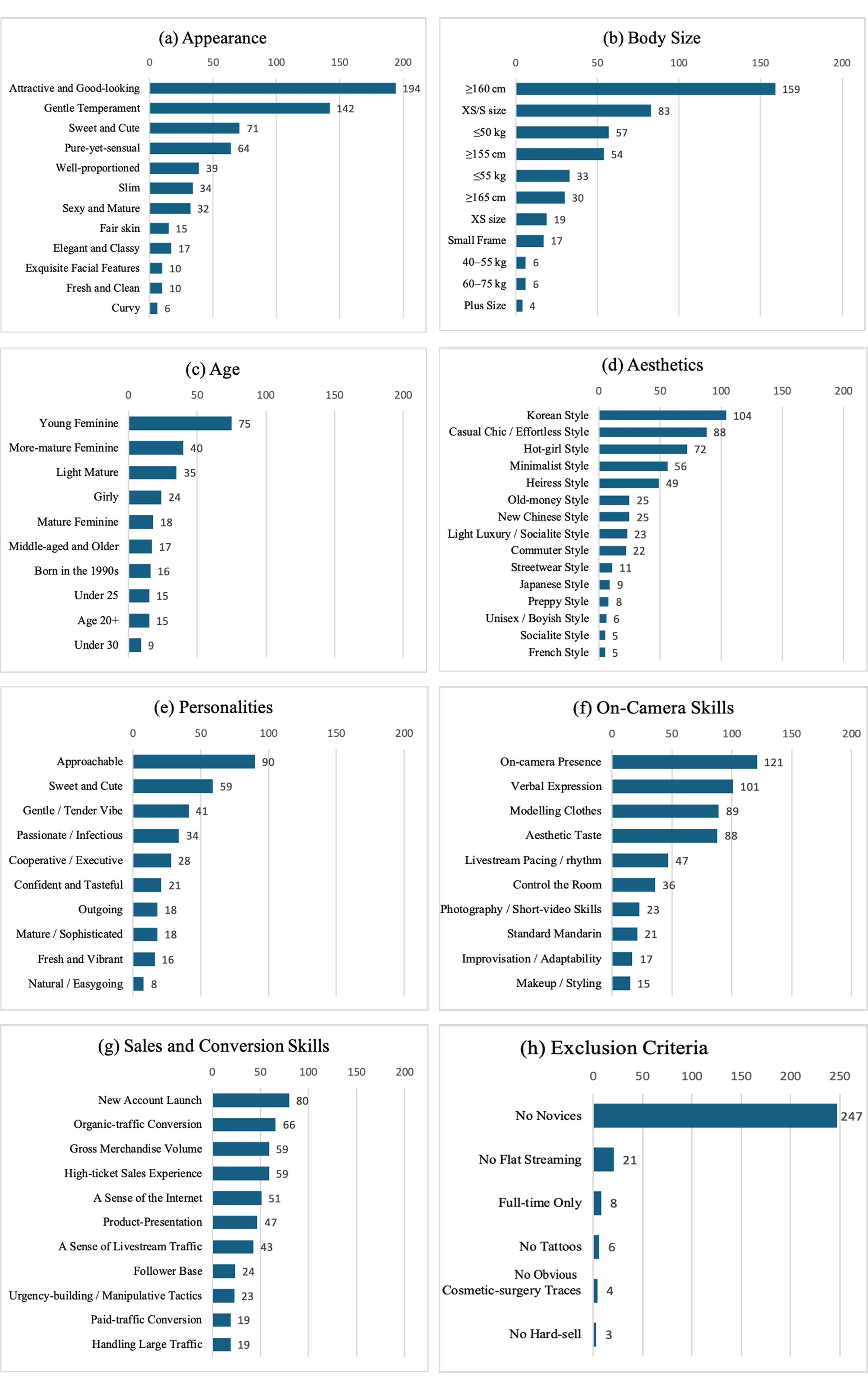}
  \caption{Recruitment requirements for women streamers across eight themes: (a) appearance, (b) body size, (c) age, (d) aesthetics, (e) personalities, (f) on-camera skills, (g) sales and conversion skills, and (h) exclusion criteria. The counts represent coded mentions.}
  \Description{A multi-panel bar chart summarizing recruitment requirements for women streamers across appearance, body size, age, aesthetics, personalities, on-camera skills, sales and conversion skills, and exclusion criteria. Each bar shows the number of coded mentions for a specific requirement.} \vspace{-10pt}
  \label{fig:recruitment-requirements}
\end{figure}

\subsection{\textbf{Appearance and Bodies}}
Appearance was the most explicitly articulated dimension of a desirable woman. Recruitment posts frequently requested women who were textit{attractive and good-looking} (194), had a \textit{gentle temperament} (142), appeared \textit{sweet and cute} (71) or \textit{pure-yet-sensual} (64). Other posts sought women described as thin, fair-skinned, or having exquisite facial features. These descriptions were often paired with precise bodily measurements. Posts requested women at least 160~cm tall (159), able to wear a size XS/S (83), or weighing no more than 50~kg (57). When age was specified, especially in women's fashion, employers also commonly favored youthfully presented women. Posts used labels such as \textit{young feminine} (75), \textit{lightly mature} (35), and \textit{girly} (24). These requirements framed the body as an occupational qualification: it was expected to appear relatively young, fair, slim, and proportionate on camera.

\subsection{\textbf{Aesthetics and Personalities}}
Bodily and facial features were emphasized as the core \textbf{aesthetic} attributes. The most frequently mentioned was \textit{Korean style} (104), often characterized by fair, luminous skin and a gentle, slim figure. Other common categories included \textit{casual chic} (88), describing an effortless yet fashionable appearance; \textit{hot-girl} (72), which emphasized a young, attractive look and a more visibly curvy or sexy body; and \textit{minimalist} (56), which resembled a clean commuter style with simple but designed clothing. \textit{heiress} (49) and \textit{old-money} (25) both conveyed a more luxurious and affluent appearance. These categories linked bodily features and styling choices to recognizable aesthetic personas that employers expected viewers to identify quickly. \textbf{Personality} requirements further specified how a desirable woman was expected to interact with viewers. The most common trait was \textit{approachable} (90), followed by \textit{sweet and cute} (59), \textit{gentle or tender} (41), and \textit{passionate or infectious} (34). These traits emphasized warmth and emotional accessibility, which helped streamers sustain conversation, build rapport, and encourage viewers to remain in the room. Other requirements, such as being \textit{cooperative and executive} (28), \textit{confident and tasteful} (21), \textit{outgoing} (18), \textit{mature and sophisticated} (18), and \textit{fresh and vibrant} (16), reflected different commercial roles and audience expectations.

\subsection{\textbf{Skills for Attention and Monetization}}
A desirable woman streamer needed to attract and sustain attention with strong on-camera skills. Recruitment posts commonly requested \textit{strong on-camera presence} (121), \textit{verbal expression} (101), the \textit{ability to model clothing well} (89), \textit{aesthetic judgment} (88), \textit{control over livestream pacing and rhythm} (47), and the \textit{ability to control the room} (36). Employers also sought improvisational ability, standard Mandarin, photography and short-video skills, and competence with makeup and styling. Streamers were expected to keep speaking, respond quickly to comments, model products, adjust their emotional intensity, and maintain enough energy to keep viewers in the room. Attention was further expected to convert into commercial outcomes with sales and conversion skills, such as gift revenue in entertainment livestreaming and sales volume in e-commerce streams. Posts requested \textit{account launch experience} (80), \textit{organic-traffic conversion} (66), \textit{demonstrated GMV (Gross Merchandise Volume) or sales performance} (59), \textit{high-ticket sales experience} (59), and an \textit{internet-ready sensibility} (51). 


\begin{figure*}[t]
    \centering
    \includegraphics[width=\textwidth]{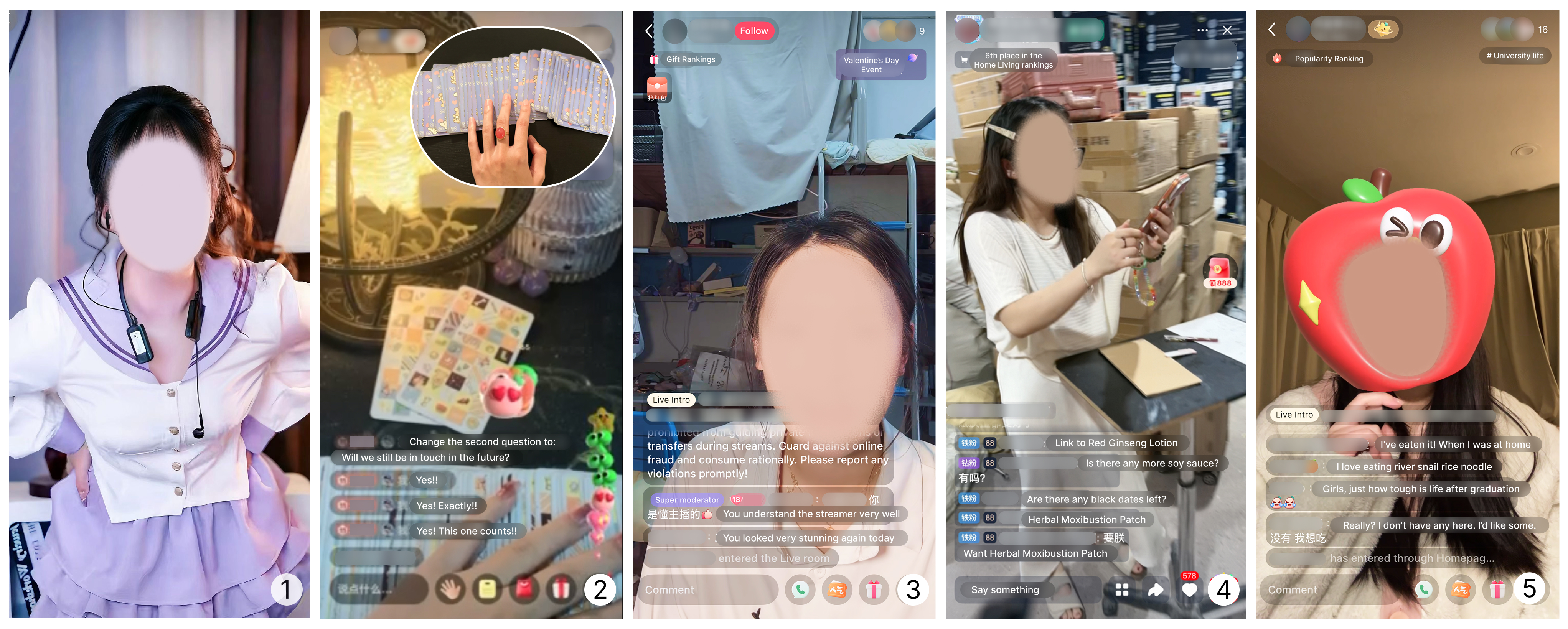} \vspace{-15pt}
    \caption{Examples of how participants assembled physical elements, filters, and backgrounds as part of their on-stream masks: (1) P15's sailor suit, pink-white lighting, low-angle camera, and rosy filter supported a pure and cute persona; (2) P2's tarot cards and gemstone rings created a mysterious, witch-like tarot-reader persona; (3) P12's dormitory background presented her as a young and energetic university student streamer; (4) P7's company warehouse setting conveyed authenticity by showing that customers would receive the products as they appeared on stream; and (5) P23's apple-head effect and home background created an authentic, relaxed, and conversational atmosphere.} \vspace{-10pt}    \Description{Five livestream screenshots showing how streamers use clothing, objects, lighting, camera angles, filters, and backgrounds. (1) P15 wears a sailor-style outfit under pink-white lighting, filmed from a low camera angle with a rosy beauty filter. (2) P2 sits behind a table with tarot cards and gemstone rings. (3) P12 livestreams from a relatively cluttered dorm room. (4) P7 livestreams from a company warehouse with products stored in the background. (5) P23 livestreams from a corner of the home using an apple-head effect.} \vspace{10pt}
        \label{fig:livestream-environments}
\end{figure*}

\section{Mask, Persona, and Livestreamers' Negotiation}
The companies or brands streamers worked with first translated the industry requirements into a more specific \textit{\textbf{persona}} for each streamer, as their expected on-stream character. Once a persona had been established, participants had to make it visually recognizable on camera. We depict this process as \textit{\textbf{masking}}. Streamers then assembled the \textit{masks} to approach these idealized \textit{personas}.

\subsection{Assembling the Mask: Constructing an On-Stream Persona}

\subsubsection{Setting the Persona}
Participants first established who they were expected to appear as on stream. For streamers working with companies, recruiters or operations staff often shaped this positioning by \textbf{assessing their appearance, voice, personality, and on-camera presence}, and matching these characteristics with particular brands or livestreaming verticals. P20 explained that women considered more \textit{attractive and good-looking} were often directed toward entertainment livestreaming, while others might be positioned for e-commerce. Entertainment companies further differentiated streamers through recognizable feminine personas such as \textit{gentle}, \textit{sweet and cute}, or \textit{pure-yet-sensual}.

This positioning also depended on the \textbf{product being sold}, particularly in e-commerce. P25 described applying to sell light-luxury shoes, in which only her legs would be livestreamed, and staff required her voice and \textit{verbal expression} to be sufficiently \textit{elegant and classy}. However, in livestreaming for another company selling affordable outdoor clothing, the case was different, as she explained:

\begin{quote}
\textit{``They described my positioning as `the kind of good daughter-in-law every mother-in-law would like.' My face is round yet slightly square, so my appearance comes across as more approachable.''} (P25)
\end{quote}

These personas were also related to how streamers already looked or behaved. Nine participants felt their assigned personas broadly aligned with their appearance, aesthetics, or personality. Four deliberately chose operations staff whose aesthetic preferences resembled their own, while P12 explained that keeping her livestream persona close to her offline self made it easier to sustain. P6 similarly noted that substantial mismatches were often resolved during recruitment, when companies selected streamers whose existing characteristics already fit the desired persona. Setting the persona hence involved matching commercial expectations with characteristics streamers could plausibly embody.

\subsubsection{Masking Through a Polished Face} 
The face was the most common site of this work. Rather than relying on beauty filters alone, streamers assembled a polished face by coordinating physical makeup, lighting, camera conditions, and digital retouching. Twenty-three participants described combining makeup with digital beauty effects, while nine reported wearing heavier makeup during livestreams because lighting, cameras, and filters softened its appearance on screen. For instance, developing P11's gentle, girl-next-door persona required simultaneous adjustments to skin tone, facial shape, makeup, contouring, filters, and lighting:

\begin{quote}
\textcolor{black}{\textit{``Our guild has a standard set of filter settings, and I usually use the naturally rosy complexion' filter because it makes me look softer and fairer. I also keep my makeup light, almost like no-makeup' makeup, without long or sharply raised eyeliner. I use contouring to make my face look smaller and its shape smoother[...] With the soft lighting, all of this makes me look gentler, like a girl next door.''} (P11)}
\end{quote}

\begin{figure*}[t]
    \centering
    \includegraphics[width=0.7\textwidth]{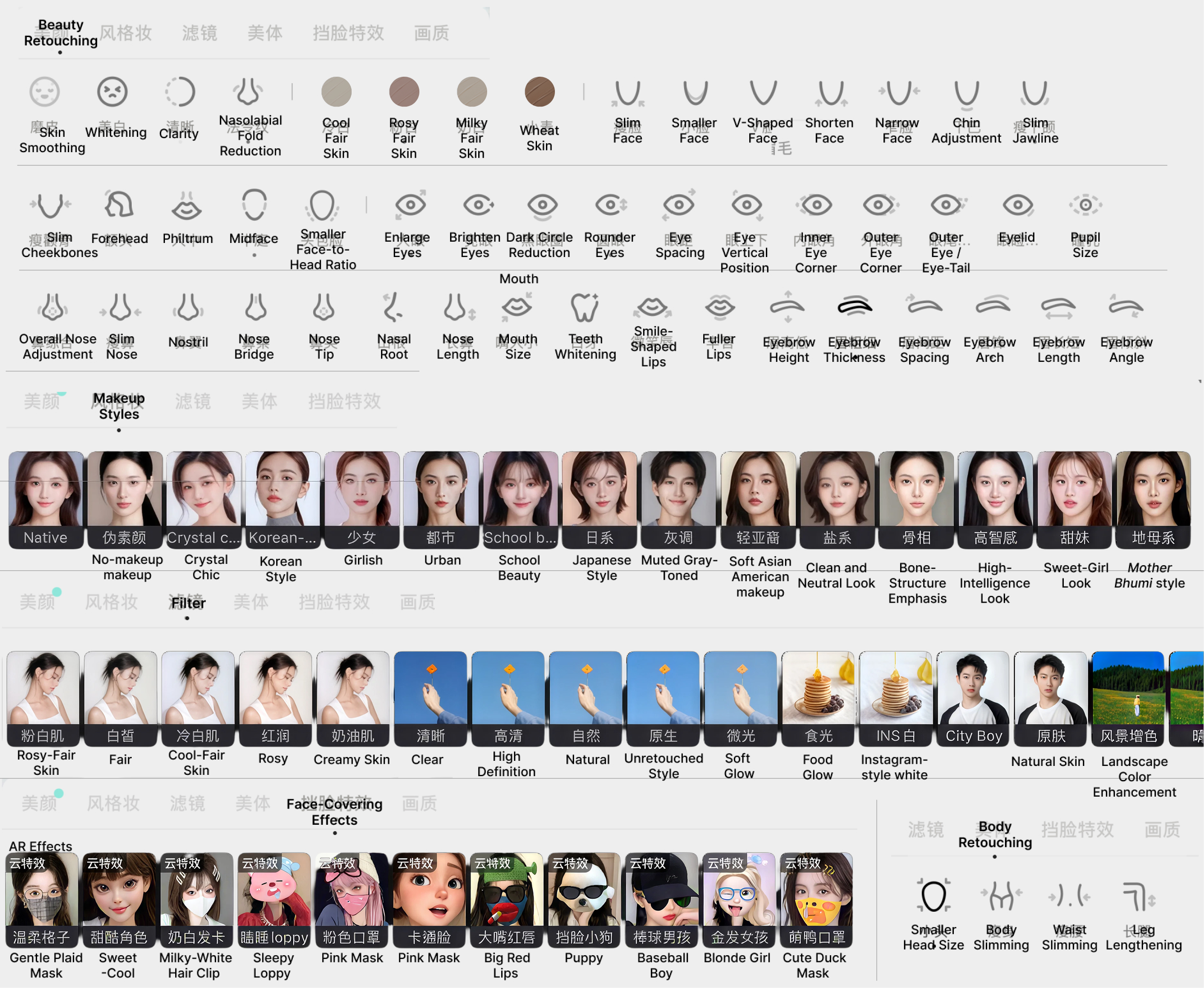}
    \vspace{-5pt}
    \caption{The beauty interface (partial) for adjusting beauty retouching, filters, and effects before going live on RedNote supported by AI technologies (translated from Mandarin). It provided granular controls over skin texture and tone, facial proportions, and other appearance features. It also offered virtual makeup effects powered by AR/AI technologies, such as the \textit{Gentle Plaid Mask}, which combines pink-toned eye makeup with a plaid face-mask effect.}
    \Description{A composite screenshot of a Chinese livestreaming beauty-retouching interface. The interface provides controls for skin texture and tone, face shape, eyes, nose, mouth, eyebrows, body reshaping, makeup-style presets, filters, and face-covering augmented-reality effects. Individual controls include skin smoothing, face slimming, eye enlargement, cheekbone slimming, philtrum and midface adjustment, nose reshaping, eyebrow adjustment, waist slimming, and leg lengthening.} 
    \label{fig:parameters}
\end{figure*}

Figure~\ref{fig:parameters} shows how this coordination was built into the \textbf{livestream interface and filters}: RedNote Live allowed streamers to manipulate skin texture and tone, facial proportions, virtual makeup, AI filters, and other visual effects. Participants did not simply maximize beautification. Twenty described appearing more attractive while remaining \textit{natural-looking} as a shared principle. They moderated individual elements or used one to compensate for another. P18, whose \textit{tuchun} persona emphasized youthfulness, purity, and naivety, avoided heavy makeup and relied on several physical lights, a reflector, and a pinkish-white skin filter. P16, who titled her stream \textit{plain-faced chatting}, used tone-up cream to make her skin appear naturally fairer while leaving some blemishes and freckles visible. A polished face therefore depended not on eliminating every perceived imperfection, but on calibrating physical and digital techniques so the result remained enhanced and plausibly natural on camera. The polished face thus became a primary surface through which the mask was assembled and made visually credible.

\subsubsection{Masking Through Body and Space}

Participants also assembled masks through bodies, clothing, objects, and the physical space visible within the camera frame. A face did not need to appear for a persona to become recognizable. P2, for example, never showed her face during tarot livestreams. She positioned the camera above a table so that viewers saw only her hands, tarot cards, and gemstone rings covered in warm yellow lighting:

\begin{quote}
    \textcolor{black}{\textit{``When I take off the mask, I am a calm and rational programmer. [...] But once I put on these rings and make the lighting warmer and darker, the gemstones stand out, and I start to look like someone who practices mysticism. Many viewers trust my expertise intuitively when they see my rings.''}}
\end{quote}

P2 called the gemstone rings \textit{``masks for my hands.''} She similarly described her warm physical lamp as a \textit{``filter''} because it changed the atmosphere of the entire camera frame. Viewers associated the rings, cards, and dim lighting with mysticism and tarot expertise. Her professional \textit{``witch-like''} persona could therefore become recognizable without showing her face.


Physical environments could extend the same \textbf{masking beyond the body}. P7 increasingly livestreamed from factories, warehouses, and sourcing locations because visible products and storage spaces supported the credibility of her e-commerce streams. Dormitories carried a different meaning. P5, P12, P15, and P23 had used university dormitories as livestream backgrounds. The ordinary furniture and relatively unpolished setting supported younger, everyday personas and made the streamer appear more accessible. The mask was therefore distributed across the body, objects, and the portion of physical space made visible to viewers.

\subsubsection{Masking Through Skilled Performance}

A recognizable mask also depended on skills such as on-camera presence, verbal expression, and the ability to control the room. Participants had to keep speaking, recognize viewers, respond quickly, manage interruptions, and sustain their persona for hours. P15 compared this work to role-playing:

\begin{quote}
\textit{``A streamer's persona is a little like a character in a role-playing game. You have to play the role you are assigned. If you choose a role that already fits who you are, it is easier to get into character.''} (P15)
\end{quote}

Product-selling streamers performed expertise through preparation, rapid presentation, and coordinated use of their bodies and surroundings. P20 described learning scripts for clothing, cosmetics, and food, including techniques for dramatizing scarcity and \textit{``immediate''} effects. Streamers might claim that only 100 units remained and would sell out once the link went live, or present a garment as immediately improving the wearer's body proportions. P26, both a streamer and MCN owner, explained that such performances were part of evaluating streamers' sales capabilities:

\begin{quote}
    \textcolor{black}{\textit{``A streamer has to believe in the product and make viewers believe in its effects. There are also many tips. For example, if I am selling a `pore-less' cushion foundation, in everyday life one pump might be enough for my whole face [...] During a livestream, I would use two pumps and apply it almost like painting a wall. Then, of course, you can't see any pores. In real life, wearing foundation that thick might look scary, but on camera, with the lighting and filters, it looks natural.''} (P26)}
\end{quote}

Here, expertise involved coordinating product use, makeup, lighting, and digital modification for the camera. Entertainment streamers developed a different repertoire for sustaining attention and interaction. Before going live, participants reviewed memes, trending topics, songs, gift-giving phrases, and information about returning viewers. Companies sometimes provided these materials, while streamers also built their own by watching larger streamers and reflecting on past broadcasts. P15, for example, filled a notebook with jokes and rhyming phrases for thanking viewers who sent gifts, helping her respond quickly while maintaining a humorous persona. Through these rehearsed performances, the mask became something streamers continually enacted rather than merely displayed.

\subsection{Fitting, Maintaining, or Breaking the Mask}

As participants continued livestreaming, the masks assembled before and during a broadcast rarely remained fixed. Participants consequently adjusted themselves to masks, adjusted masks to changing conditions, maintained them with other people, sometimes broke them, and set boundaries for them.

\subsubsection{Fitting the Mask to Self, Market, and Platform}

Fitting a mask involved finding a workable distance between the desirable persona and the person performing it. Some participants could amplify characteristics they already had, while others found company-assigned personas harder to sustain. P3, for example, struggled with a \textit{``sexy and mature''} role in group livestreaming because it required sexually suggestive movements (\textit{cabian}) that felt distant from her quiet personality. P11 similarly described the effort of maintaining a persona that differed from her offline self:

\begin{quote}
\textcolor{black}{\textit{``My company assigned me a gentle persona, but I am actually quite lively in everyday life. So on-stream, I have to act gentler and more composed. I soften my voice and sometimes speak as if I have less energy. I also have to control my expressions and make my movements smaller.''} (P11)}
\end{quote}

Although she could perform the role, maintaining it for hours required extra effort. Participants therefore preferred personas that remained close enough to themselves. This tension also shaped how far streamers were willing to modify their appearance. Operations staff sometimes preset beauty parameters, and seven participants described disagreements over these adjustments. Four resisted excessive face slimming, eye enlargement, or whitening when they felt that the resulting face no longer resembled them. P3 described how operations staff could reproduce similar feminine templates across different streamers:

\begin{quote}
\textcolor{black}{\textit{``They may prefer thick eyebrows, big eyes, and that influencer (wanghong') face. They think a more youthful, sweet, vulnerable girl is more attractive to men. They like that kind of look, so when they adjust your face, they try to make your facial contours rounder and softer. But we streamers still want to keep some of our own features. If I adjust the settings myself, I turn them down. I would not max everything out. Otherwise, some livestream rooms make everyone look almost like they are wearing a fake mask.''} (P3)}
\end{quote}

At the same time, the mainstream \textbf{aesthetic standards} against which masks were fitted were not produced only by companies. Participants repeatedly encountered preferences for fairer skin, thinner faces and bodies, and more youthful appearances across livestreaming platforms and gradually incorporated some of these standards into their own judgments. P12 said she would \textit{``subconsciously''} increase whitening and skin smoothing because these adjustments had become part of mainstream livestream aesthetics, even though she preferred a relatively natural appearance. P18 set whitening to its maximum level because streamers in the same genre appeared especially \textit{``fair and luminous'',} even though some viewers told her that she looked too white. As P15 shared, these preferences could also change over time:

\begin{quote}
\textcolor{black}{\textit{``There was a time when I was really fixated on having a small, short face and an overall cute, anime-like look. I used pretty strong beauty filters then. If I couldn’t adjust them to a look I was satisfied with, I didn’t even want to start streaming. It actually made me very anxious about my appearance, because I would look in the mirror before and after streaming and feel completely caught up in appearance anxiety.''} (P15)}
\end{quote}

For product-selling streamers, the mask also had to support the product’s credibility. After a viewer commented that she looked too heavy, P20 began using waist-slimming effects when selling slimming patches to make the product appear more effective. In contrast, P5 and P10 reduced beauty effects and filter use when selling fabric because these effects could distort the product’s color and texture:

\begin{quote}
    \textcolor{black}{\textit{``In Douyin videos, I like using skin smoothing to remove wrinkles. But I can’t do that when livestreaming fabric. Viewers need to see the texture clearly in high definition. [...] Something that is originally pink might look light pink after a filter brightens it. We only use white lights and try not to change the fabric’s color. If customers receive it and see a color difference, they might leave a bad review or return it.''}(P5)}
\end{quote}

The mask also had to abide by \textbf{platform-appropriate behaviors}. All 26 participants described reduced traffic, livestream suspensions, mistaken moderation, or account bans. Participants learned platform restrictions through company training, operations staff, and trial and error, adjusting both speech and performance accordingly. E-commerce streamers avoided absolute claims such as \textit{100\% effective} or \textit{the best} while others substituted coded expressions for restricted terms. Adding someone on WeChat could become \textit{adding Paopao}, while asking viewers to recharge money could be conveyed through a \textit{C} hand gesture. Such substitutions allowed experienced viewers to understand what was being communicated while making the performance less legible to moderation systems. However, \textbf{platform standards are vague and often difficult to predict}. P15 described avoiding numerous kinship terms and suggestive expressions for years, while still encountering seemingly inconsistent clothing moderation:

\begin{quote}
    \textcolor{black}{\textit{``I feel that Douyin does not really have a clear standard. I have worn a camisole or a silly costume headpiece with no problem at all, but once I wore a more stylish shirt with a cutout at the shoulder and immediately started receiving warnings for `unusual clothing.' ''} (P15)}
\end{quote}

Since tarot content was classified as \textit{`feudal superstition,''} P2 tried various ways to bypass platform moderation, compromising her witch-like persona in the process. For example, she wrote slogans such as \textit{`Love science''} on the whiteboard in her livestream room to reduce the risk of being flagged:

\begin{quote}
    \textcolor{black}{\textit{``As long as the cards appeared on camera and viewers could see them, the risk was very high. I had already titled my livestream `Poker Card Game' [...] RedNote gives you a whiteboard where you can write notices, so I wrote things like `Love science' and `Long live Chairman Mao' to avoid getting banned. During my most popular stream, there were several hundred people watching, and suddenly I was banned. The whole livestream room just disappeared.''} (P2)}
\end{quote}

She eventually stopped livestreaming after two of her accounts were banned. Fitting the mask was thus never complete: streamers continually adjusted it to remain close enough to themselves to sustain, credible enough for the market to accept, and legible enough to viewers while remaining visible within platform systems.

\subsubsection{Maintaining the Mask with Audience}
Maintaining a mask became a relational task once viewers began interacting with it. A streamer could prepare a \textit{gentle}, \textit{approachable}, or humorous persona before going live, but audiences continually tested whether that persona remained coherent and also participated in shaping it through their interactions. Room moderators (\textit{fangguan}) played a particularly important role in this process. They helped streamers maintain personas by performing actions that the streamer did not want to make visible. P15 explained that directly arguing with hostile viewers could contradict the persona she was performing:

\begin{quote}
    \textcolor{black}{\textit{``I have my persona in the livestream. My persona may not allow me to say certain things. If someone curses at me, I cannot directly fight back. [...] But I can privately tell the room moderator, or the moderator will see something inappropriate and deal with it for me.''} (P15)}
\end{quote}

\textbf{Moderators' capacity} included reporting viewers, responding to hostile comments, or posting enough messages to push unwanted comments out of view. This allowed streamers to preserve a warm or controlled public presentation while disciplinary work happened around them. Because moderators were often long-term fans, this also created relational labor. Three participants described reassuring moderators who became jealous when streamers paid more attention to new high-spending viewers. \textbf{Audience participation} also helped maintain technologically mediated performances. P15 described AI-assisted singing systems that generated requested songs using a prerecorded version of a streamer's voice. Regular viewers often understood how the system worked but continued paying to request songs. She described this as \textit{xinzhao buxuan}, or being mutually understood without being stated. Comments calling the performance \textit{`AI singing''} or \textit{`fake singing''} could be blocked, making the mask dependent on a shared convention between streamer and audience. Entertainment streamers also carried these personas into private messages and social-media accounts, especially with viewers who regularly sent large \textbf{virtual gifts}. Five participants added these high-spending viewers, commonly called \textit{dage} (literally big brothers), on personal WeChat accounts. P11 used a separate work account to post photos and sustain the same \textit{gentle} image outside the livestream. Sustaining a recognizable persona could therefore continue after the public broadcast ended.

\subsubsection{Breaking the Mask: Creating Contrast and Managing Breakdowns}

Breaking a mask did not always mean abandoning the persona behind it. Personas such as \textit{gentle}, \textit{sweet and cute}, or \textit{elegant} remained relatively abstract, while masks were their concrete on-stream forms assembled through appearance, technology, space, voice, and performance. Participants sometimes broke or replaced these masks strategically to create contrast and novelty. At other times, technical failures or livestream interactions disrupted them unexpectedly.

First, \textbf{showing viewers an unexpected side of themselves} was a deliberate way to break a polished mask. P14, for example, often appeared as a polished and conventionally attractive entertainment streamer, then abruptly created what she called \textit{fancha}, or contrast, by making exaggerated expressions, acting deliberately silly, or imitating cartoon characters. She described this as an \textit{``abstract''} style of humor:

\begin{quote}
    \textcolor{black}{\textit{``Viewers like contrast. A quiet, pretty streamer suddenly starts acting silly. For example, I might suddenly imitate SpongeBob's voice. They will say, `If you're going to act, then act properly. Why did your real self suddenly come out?' And I say, `Sorry, I'll switch back now.' ''}}
\end{quote}

In this case, regular viewers recognized these switches and incorporated them into their interactions with the stream. The break itself could therefore become another recognizable layer of the persona. P14 used a similar strategy in her ``Gacha pull and blind boxes'' livestream. Across streams, she shifted among \textit{mature and sexy}, \textit{pure-yet-sensual}, \textit{sweet and cute}, and \textit{gentle} masks. She compared these shifts to opening a blind box, as viewers could enter the livestream to see which version of her would appear that day. Breaking or replacing one mask could thus create novelty while keeping the persona more recognizable.

\textbf{Technical failure }was another reason for breaking the masks. Twenty-four participants described beauty effects temporarily losing track of their faces or bodies. Face slimming could stop when an object crossed the face, making the face suddenly change size, while skin smoothing could briefly fail and reveal blemishes. Some streamers treated these failures as routine. P15 said that large movements often caused beauty retouching to lose track of her face, but regular viewers were already used to it. Others experienced greater appearance anxiety. P20 worried that viewers or competing streamers might screenshot an unfiltered moment and circulate it. These failures further changed how streamers used beauty technologies. P25 stopped using the leg-lengthening effect because it often disappeared when she moved. She explained:

\begin{quote}
    \textcolor{black}{\textit{``When I used the leg-lengthening effect, it would disappear when I stepped forward, and the background would suddenly shrink. People could obviously tell that I had lengthened my legs. It looked very unreal and made it hard for people to trust me. If the background keeps moving whenever you move, people will think everything is fake. I want the women watching me to feel that I am as real as possible, so now I do not use leg-lengthening effects anymore.''}(P25)}
\end{quote}

For P25, the glitch exposed the manipulation itself and threatened the sense of authenticity she wanted to maintain with viewers. She therefore preferred a less modified body that could remain visually stable throughout the stream.

Another form of mask breaking happens through \textbf{planned unexpected behavior as a performance} by livestreamers, particularly in entertainment. During \textit{PK} sessions, two or more streamers competed for points generated through viewers' gifts and likes, and the losing streamer could be asked to complete a punishment. Some punishments involved \textit{``turning off beauty retouching''} or \textit{``removing makeup.''} Participants prepared for this possibility by keeping beauty parameters relatively natural and relying on physical makeup and lighting as additional layers of the mask. Loyal viewers could also send more gifts to help the streamer avoid losing, while streamers sometimes negotiated alternative punishments with one another so that makeup removal did not actually occur. Even when breaking the mask was built into the game, streamers and audiences worked together to control how much of it would be exposed.

\subsubsection{Protecting the Boundaries}
Beyond managing how a mask appeared, participants also negotiated where that mask should end. They used stage names, selective camera framing, separate accounts, and changes in self-presentation to keep parts of their identity and everyday lives outside the livestream. These boundaries became especially important when viewers sought greater access to the person behind the persona.

Participants first managed \textbf{boundaries around names and physical spaces}. P25 consistently used the stage name \textit{``Bamboo''} across different livestreaming jobs. P15 similarly used an online name and felt awkward when viewers discovered her legal name through Douyin's identity information. Participants streaming from bedrooms or dormitories also positioned cameras so that only selected parts of the room remained visible. P12 described:

\begin{quote}
\textcolor{black}{\textit{``I care a lot about other people's privacy. Now I only stream during summer break when my roommates are away. [...] One time, my roommate suddenly came back, so I completely blocked the camera. I would not let her appear on screen at all. I just used my face to cover the frame. My roommate does not know that I livestream because I think people can have negative views of the livestreaming industry.''} (P12)}
\end{quote}

P16 similarly stopped streaming to \textbf{protect the privacy} of her roommate. When viewers offered to send food or drinks, P7 and P9 provided company addresses or pickup locations instead of their homes. For P15, these precautions became especially important after viewers inferred her offline location from a dormitory livestream. She recalled:

\begin{quote}
\textcolor{black}{\textit{``I was livestreaming from my dorm, and the platform had a `nearby search' function. People in the same city could come across my stream and know I was broadcasting from a dorm. Dorms all look quite similar, so they could identify my university quite precisely. I once had a very obsessive fan wait for me at the campus gate. He was really obsessive[...] I had to call the police.''}(P15)}
\end{quote}

The boundary of the mask also extended to people in streamers' offline lives. Because women livestreamers could face gendered stigma, three participants \textbf{concealed their work from family members or acquaintances}. P3 did not tell her family that she worked as an entertainment streamer because she expected older relatives to view it as an improper occupation. P13 was considering developing a couples account but hesitated to bring her husband on camera:

\begin{quote}
\textcolor{black}{\textit{``Right now, I probably would not let him appear on camera. I would want him to put on some makeup or fix his hair after work, but that is not very realistic because he is already exhausted [...] I want both of us to appear in a good state. There are all kinds of people watching livestreams, they look at you and your partner and start judging the relationship[...] If one person does not look attractive, they start asking, `Why is this woman with this man?' or `Why is this man with this woman?' I do not want viewers judging my family.''}(P13)}
\end{quote}

For P13, the \textbf{boundary of the mask was spaced by the care} for her husband. Keeping him outside the frame protected him from viewers' scrutiny and from being judged through an unmasked appearance. P15 encountered a different conflict between her online persona and offline relationships. In a voice room, operations staff positioned her voice as \textit{sexy and mature} and suggested that she post back-view or partially hidden photographs of women with a matching appearance so viewers could more easily imagine this persona. P15 instead posted her own photographs. Her offline friends also followed the account, and she did not want them to see photographs of another woman and become confused about who she was presenting herself to be. Here, the mask had to remain workable across livestream audiences and people who already knew her offline.

\textbf{Gendered audience} relationships created another boundary around the intimacy implied by a persona. P6, P20, and P25 encountered \textbf{sexually abusive} comments while selling menswear or underwear, while P22 received explicit questions when hosting men's health livestreams. Such interactions were especially common among entertainment streamers, who described high-spending male viewers treating virtual gifts as grounds for romantic attention or offline access. Participants sometimes called these gifts \textit{aimei piao}, or \textit{``flirtatious gifts.''} P14 explained that the companies and guilds she had worked with explicitly prohibited this practice:

\begin{quote}
\textcolor{black}{\textit{``As a female streamer, you are actually quite afraid of `aimei piao.' A viewer might flirt with the streamer and then send gifts worth far more than the performance itself. At our company, and at several companies and guilds I signed with before, there was always a `red line' against flirting. We were prohibited from making money through flirtation because it could cause a lot of trouble and make the atmosphere in the livestream room uncomfortable.'' (P22)}}
\end{quote}

Though companies treated flirtation as a red line, the support they could provide was limited, and streamers often had to handle \textbf{uninvited sexual or romantic advances from the audience on the spot}. P15, for example, avoided by exaggerating the transactional side of the interaction when viewers repeatedly asked whether she would date them. She described picking up a nearby object as if it were a phone and pretending to call her mother:

\begin{quote}
\textcolor{black}{\textit{``If he keeps asking, `Can we date?' I will pick up whatever is next to me and pretend it is a phone. I will say, `Hello, Mom? Someone here wants to date me. What? The bride price has to be a car?' [referring to an expensive virtual gift]. Most of them leave as soon as I directly ask for gifts.''}(P15)}
\end{quote}

Other streamers \textbf{deliberately weakened the attractive or intimate mask} when viewers pushed these boundaries. When a high-spending viewer wanted to meet P16 offline, she told him that she only looked attractive because of beauty filters and was not good-looking in person. P19 sometimes told male viewers that she was a lesbian to discourage romantic expectations. In these cases, partially breaking the desirable mask helped streamers reduce the intimacy that viewers attached to it and re-establish limits around offline access.

Participants also described bodily and emotional limits to how long a mask could be maintained. Livestreaming required long hours, continuous speaking, and constant attention to traffic, sales, gifts, and audience reactions. P7, who had livestreamed on Taobao for nine years, recalled once describing a single shirt continuously for 40 minutes, down to details such as individual threads. She described reaching a state in which her own sense of self seemed to disappear into the performance. P11, a newer streamer, was repeatedly assigned overnight shifts that extended into the early morning:

\begin{quote}
    \textcolor{black}{\textit{``Operators schedule me for overnight shifts until three, four, five, or even six in the morning. They said these hours could reach a specific group of target users. When I streamed until five or six, I could feel my heart beating very fast. I thought, `My body cannot keep doing this.' [...] I was also scared to go home alone at three or four in the morning, so I would wait there (in company) until around six or six-thirty, when it got light, before leaving.''} (P11)}
\end{quote}

For P11, maintaining the livestreaming performance became constrained by both physical strain and the safety risks of overnight work. After experiencing severe physical stress, she eventually left the job. Similarly, for P26, accumulated exhaustion eventually broke an otherwise stable \textit{elegant and classic} presentation. During a \textit{Double Eleven Shopping Festival} in Taobao, she sometimes livestreamed for 15 or 16 hours while sleeping only about four hours a day. After many days of this schedule, a viewer's sexually degrading comment became the \textbf{breaking point}. P26 recalled:

\begin{quote}
\textit{``My livestreaming state had always been very stable. But there was one time when I exploded. [...] I had been sleeping only about four hours every day and sometimes livestreaming for fifteen or sixteen hours. My body was already at its limit. Then someone said something like, `Selling your body is selling, and selling clothes is also selling.' I just exploded. Everything had built up to that point. I told operation staff, `Let him speak. Do not mute him. I want to confront him directly.' '' (P26)}
\end{quote}

Here, breaking the mask was no longer a planned strategy for managing an audience boundary. The physical and emotional labor required to maintain P26's controlled persona had exceeded what she could continue to absorb. Her response exposed a boundary that the \textit{``elegant and classic''} mask had previously concealed.

Eight participants responded to these demands by \textbf{reducing masking itself}. P23 deliberately avoided beauty retouching and facial filters, using only occasional playful effects that did not alter her facial features:

\begin{quote}
    \textcolor{black}{\textit{``My favorite RedNote effect is the apple headpiece. Some of the others feel ``too much''. The apple looks more natural because it only adds a headpiece and does not change my face at all. I personally dislike anything that changes my facial features. There is already so much that is fake on the internet. I do not want to add more artificial things to how I appear on screen. I want to stay as real as possible.''} (P23)}
\end{quote}

She did not join a guild and explicitly rejected the effort required to maintain the polished and beautiful presentation commonly expected of women streamers. Although her traffic remained relatively modest, she described gaining a stable group of women viewers who valued this presentation. P23's practice shows that protecting the boundary of the mask could also mean refusing to expand it further.
 
\section{Discussion}
Our study shows that masking is more than a cosmetic platform functionality. It is a political and sociotechnical negotiation over who can shape a streamer's visible self, which femininities become recognizable and valuable, how bodies and spaces become visible, and when these demands can be modified or refused. We examine these politics through assemblage, visibility, breakdown, resistance, and refusal, opening new directions for HCI theory and design.

\subsection{Implications to Design}
Our findings identify three problems requiring immediate design attention: AI beauty tools that transform appearance without protecting streamer-defined features, organizational editing without clear accountability, and opaque AI moderation that makes visibility difficult to contest.

\subsubsection{Preserving Streamer-Defined Appearance in AI Filters}
AI beauty interfaces make transformation easy but provide little support for preserving features streamers do not want optimized. P3 reduced the company-preferred ``wanghong'' face to retain her own characteristics, while P15 described anxiety over appearing insufficiently small-faced and youthful (6.2.1). Computational beautification can normalize aesthetic standards \cite{riccio2024mirror,riccio2022openfilter,LIAN2023107793}, while recent studies of AI-driven filters show normative transformation and opacity around how models act on users' faces \cite{doh2025bold,west2025impressively}. We propose \textit{preservation-aware filters}: streamers could lock facial contours, skin texture, body proportions, or product colors, set modification ranges, and save masks around protected features. Personalized moderation similarly lets users configure filter categories and intensity \cite{heung2025ignorance}. Applied to beautification, this would make AI adapt within streamer-defined limits rather than users adapting to platform-defined ideals.

\subsubsection{Making Mask Editing Contestable Across Power Relations}
Streamers could adjust masks while lacking equal power over who defined those settings. Operations staff selected personas, altered beauty parameters and lighting, and evaluated performance, reflecting organizational control in livestreaming MCNs and guilds \cite{xiao2026algorithmic,liu2023zhibo}. P3's reduction of imposed settings shows why additional sliders alone cannot resolve this imbalance. We propose \textit{role-aware mask provenance}: streamer-owned baselines, proposed organizational changes, parameter-level acceptance or rejection, editing records, and restoration of preferred configurations. \textit{Gaze and Glow} shows how exposing invisible social-media editing can support reflection and user-controlled visibility \cite{Hong2025gaze}; contestable-AI design similarly calls for control, explanation, intervention, and scrutiny \cite{alfrink2023contestable}. Contestability should therefore extend beyond AI outputs to the organizational actors configuring them, making power over women's mediated appearance visible and negotiable rather than silently encoded in presets.

\subsubsection{Making AI Moderation Legible and Appealable}
Participants often could not determine what masking platforms would permit. Subsection 6.2.1 shows P15 receiving inconsistent clothing warnings, P2 modifying her tarot persona after ``feudal superstition'' classifications, and P23 encountering an apparently erroneous fortune-telling classification. Platforms could instead identify the phrase, gesture, visual region, or clothing feature triggering intervention, show the applicable rule and system confidence, and provide human review; rehearsal modes could test masks before broadcasting. Contestable moderation emphasizes representation, communication, and opportunities to influence decisions \cite{vaccaro2021contestability}, while post-removal explanation research shows the value of sanction rationales \cite{jhaver2024bystanders}. Personalized moderation also foregrounds configurable AI filtering and heterogeneous tolerances \cite{heung2025ignorance}. Moderation should thus become an inspectable and contestable relationship, shifting interpretive burden from workers toward platforms and making them accountable for the visibility conditions they produce.

\subsection{Masking Through an Assemblage Lens}
Masking is not located in any single technology, body, or act of self-presentation. It emerges through relations among streamers, technologies, material environments, organizations, audiences, and platform systems. Assemblage theory helps explain how these elements are brought together, stabilized, disrupted, and reconfigured, and how power enters masking through the relations among them rather than through any single filter or actor.

\subsubsection{Masking as a Sociotechnical Assemblage}
We extend Deleuze and Guattari's concept of assemblage into AI-mediated self-presentation by showing that a livestreamed persona emerges through a provisional arrangement of heterogeneous material, technical, social, and expressive elements \cite{felix1987thousand,delanda2019new,holland2005gilles,muller2015assemblages}. This extends design and HCI accounts of sociotechnical assemblages, where people, tools, practices, and materialities jointly make action possible \cite{beaubois2015design,willis2019m}, and work on algorithmic subjectivities, where subjects emerge through relations among humans, algorithms, institutions, and classifications \cite{baumer2024algorithmic}. In Section~6.1, P2's ``witch-like'' tarot persona emerged through rings, cards, hand positioning, lighting, framing, speech, and platform mediation rather than her face alone; others coordinated makeup, filters, clothing, rooms, products, and performance. We therefore contribute \textit{masking assemblage} as self-presentation whose capacities emerge relationally across bodily, computational, spatial, organizational, and performative components \cite{baumer2024algorithmic,chu2025hci,delanda2019new,beaubois2015design}.

\subsubsection{Masking Through Stabilization and Reconfiguration}
We extend assemblage accounts of territorialization, deterritorialization, and reterritorialization by showing how they operate through the stabilization and disruption of gendered, technologically mediated personas \cite{felix1987thousand,delanda2019new,holland2005gilles,muller2015assemblages}. Assemblage scholarship emphasizes temporary stability and recomposition \cite{holland2005gilles,muller2015assemblages,beaubois2015design,willis2019m}; HCI similarly shows algorithmic subjectivities as constituted through changing sociotechnical relations \cite{baumer2024algorithmic}. Section~6.2 makes this visible: operations staff stabilized masks through assigned personas, beauty presets, and metrics, while audiences and moderation reinforced recognizable forms. P3 reduced the standardized ``wanghong'' face, P25 abandoned unstable leg-lengthening, P14 deliberately broke a polished mask, and P23 refused beauty retouching. These disruptions did not necessarily constitute escape: P14's contrast became another recognizable performance. Masking agency is therefore neither simple compliance nor resistance, but the capacity to destabilize, preserve, and recompose relations within a commercial sociotechnical assemblage \cite{muller2015assemblages,willis2019m,baumer2024algorithmic,chu2025hci}.

\subsubsection{Designing for Reconfigurable Mask Assemblages}
We extend design-oriented assemblage theory by conceptualizing streamer agency as the capacity to reconfigure relations within a mask assemblage, rather than merely adding interface controls \cite{felix1987thousand,delanda2019new,beaubois2015design,willis2019m}. Design scholarship locates artifact functionality within wider sociotechnical arrangements \cite{beaubois2015design}, while HCI emphasizes relationality, reconfigurability, and alternative arrangements \cite{willis2019m,loscri2024road}. Control over masking was distributed across streamers, operations staff, viewers, moderators, platform rules, AI beauty systems, and physical environments. In Sections~6.2.1--6.2.4, participants reconfigured these relations by rejecting presets, changing visibility boundaries, delegating hostile audience management, reducing filters, or withdrawing. This extends feminist HCI and non-use scholarship by locating refusal within the relations that make particular forms of visibility possible \cite{Feministhci,satchell2009beyond}. Designing for masking should therefore support provenance, negotiable permissions, reversible configurations, protected boundaries, rehearsal spaces, and partial disengagement, enabling streamers to reorganize who and what can shape their visible selves \cite{loscri2024road,Feministhci,satchell2009beyond,baumer2024algorithmic}.

\subsection{The Politics of Visibility and Boundary Work}
Beyond modifying appearance, masking organized who and what could become visible across spatial, relational, and gendered boundaries, revealing how visibility was negotiated and unequally valued.

\subsubsection{Spatial and Relational Boundaries of Visibility}
Masking reorganized visibility across domestic, public, and commercial space. Participants used framing, lighting, props, stage names, separate accounts, and interruptions to decide what entered the stream. Building on work showing that space is produced through embodied practice and technological appropriation \cite{dourish2006respace}, digital commerce reshapes domestic environments \cite{Mim2022fcommerce}, and camera placement and cohabitants structure livestream visibility \cite{tran2024room,Wu2022concerned}, we show that the camera frame became part of the mask itself. Privacy was therefore collective and relational: livestreaming can encourage disclosure \cite{Wu2022concerned,Ahmed2019privacy}, while protecting roommates, partners, and families became care enacted through visibility boundaries \cite{vallgarda2025attuning}. Commercial intimacy \cite{ye2023ambivalent} similarly concentrated physical, technical, emotional, and relational labor \cite{zhang2019livestreaming,ye2023ambivalent,zou2018producing}, intensified by metrics and algorithmic evaluation \cite{xiao2026algorithmic,liang2025pedagogy,Qiao2026live,wang2024harnessing}. Yet gendered stereotypes can sexualize streamers while obscuring these skills \cite{zhang2019nuzhubo,ruberg2019titty,tang2022dare}, and polished systems may under-recognize ``raw'' authenticity \cite{tang2022dare}. Masking therefore shifts privacy from an individual problem toward how platforms distribute the costs of visibility across relational environments.

\subsubsection{Gendered Recognition and the Politics of Legibility}
Masking also reveals recognition as a negotiated achievement shaped by culturally available signs and unequal evaluation. In \textit{Nuo} rituals and Chinese opera, masks, costume, movement, color, and performance make roles recognizable through shared conventions \cite{li2024nuo,shen2022bodily,yang2025nuomask,tu2017cultural}; participants similarly assembled \textit{gentle}, \textit{elegant}, or \textit{sweet and cute} personas across bodies, technologies, spaces, and performance. Extending Jung's \textit{persona} as a compromise between individual and social expectation \cite{jung1966two}, we show this compromise becoming occupational and infrastructural as companies, audiences, recommendation, and moderation systems shape recognizable and permissible personas \cite{xiao2026algorithmic,liang2025pedagogy,Qiao2026live}. The mask thus becomes an emergent \textit{boundary object} \cite{bala2024stories} within symbolic boundaries that distribute value unevenly \cite{krstic2022symbolic,lamont2002study}. Recruitment standards and computational beautification normalize fair, thin, youthful femininity \cite{riccio2024mirror,riccio2022openfilter,LIAN2023107793}, while postfeminist optimization \cite{postfeminist} and ``naturalness'' \cite{you2024competitive} can reproduce the same field. Streamers nevertheless redrew boundaries around professionalism, \textit{cabian}, intimacy, and identity \cite{zhang2019nuzhubo,zhang2019livestreaming,tang2022dare,ye2023ambivalent}, even when these distinctions could stigmatize other women \cite{krstic2022symbolic,goffman1963stigma}. Feminist HCI \cite{Feministhci} sharpens the contradiction: supporting recognition without requiring legibility through the same normative categories that constrain people remains a central challenge.

\subsection{Breaking, Resistance, and Refusal}
The politics of masking become especially visible when a mask no longer holds. Breaks ranged from technical failures to strategic departures, resistance, and refusal, exposing the labor of sustaining a persona and the unequal costs of withdrawal.

\subsubsection{Breakdown as Productive Exposure}
Breaking reveals that a livestream persona is continuously maintained rather than fixed. HCI treats breakdown as productive: broken technologies can expose backgrounded relations with artifacts and create opportunities for repurposing, alternative values, and creativity \cite{jackson2014breakdown}. Generative theories similarly treat breakdown as a site of learning that seeds future practices \cite{beaudouinlafon2021generative}, while glitch research shows how disruption can generate new practices, aesthetics, and values \cite{garrett2025glitch}. In our findings, deliberate contrast turned departure from an established mask into performance, while filter failures exposed the fragility of a polished body and reshaped later practices. Breaks thus exposed not a more ``authentic'' self beneath the mask, but the infrastructures and dependencies that ordinarily make a mediated self appear seamless.

\subsubsection{Breaking as Soft Resistance}
Breaking could also express partial agency within organizational constraints. Wong's \textit{soft resistance} describes how workers alter practices from positions of limited power while remaining within many of the same logics they seek to change \cite{wong2021soft}. Streamers similarly lowered company-defined beauty settings, modified assigned personas, or performed them differently while remaining in commercial roles. Breaking created room to renegotiate how closely industry-defined forms governed appearance without abandoning the mask, yet audience expectations, organizational evaluation, platform metrics, and income still constrained that agency. Resistance within masking was therefore incremental and relational rather than simply oppositional. More critically, agency cannot be measured only by whether users can alter a system, but by how much deviation its economic and organizational arrangements tolerate before participation itself becomes costly.

\subsubsection{When Breaking Becomes Refusal}
Some participants moved from modifying the mask to deciding where masking itself should stop, disabling body reshaping, reducing beauty retouching, weakening attractive personas when intimacy crossed boundaries, or withdrawing when demands became unsustainable. Technology non-use provides one entry point \cite{nonuse}, but Baumer and Khovanskaya caution that use/non-use framings can obscure the social relations and changing subjectivities through which disengagement becomes possible \cite{Remantling}. Refusal scholarship instead treats saying no as a situated, generative practice that reshapes relationships with sociotechnical systems \cite{zong2024refusal,pereira2021refusing}; Zong and Matias foreground autonomy, time, power, and cost \cite{zong2024refusal}. Turning off a beauty effect may be simple while jeopardizing visibility, income, audience response, or workplace evaluation. Refusal therefore marks a deeper break with the conditions that make particular appearances valuable and expected. Because resistance can be absorbed into platforms without altering underlying power relations \cite{ganesh2022resistance}, the politics of breaking ultimately concern who can afford to refuse the mask, under what conditions, and with what consequences.
\section{Limitations and Future Work}

Our study focused primarily on lower-tier (\textit{dibu}) Chinese women livestreamers, whose experiences may differ from those of middle-tier (\textit{yaobu}) or top-tier (\textit{toubu}) streamers working with larger audiences, stronger teams, and greater bargaining power. Our participants were also primarily heterosexual women, leaving room for future work to examine how masking is shaped by sexuality, gender identity, class, age, region, and other intersecting social positions. Our recruitment-post analysis relied on WeChat groups. Future studies could combine multiple recruitment platforms and industry data sources to capture a broader livestreaming labor ecology. Our study also approached livestreaming through observation and interviews. Several participants suggested that researchers could livestream themselves for an extended period to gain an insider perspective on masking. Such self-ethnographic or longer-term engagement may reveal practices that are difficult to articulate in observation and interviews. Given these boundaries, we refrain from generalizing our findings beyond the studied setting and instead emphasize the value of in-depth empirical work for revealing how masking is negotiated in practice. We hope to deepen our findings and contributions through more diverse participants, broader data sources, and extended engagement with livestreaming communities in the future.

\section{Conclusion}
We examine how Chinese women livestreamers use masking to negotiate industry requirements and their own agendas for visibility, identity, privacy, and safety. Our content analysis of recruitment posts, livestream observations, and interviews found that industry-defined ideals of appearance, femininity, and performance shape recognizable masks before women go live. Streamers assembled these masks through beauty technologies, makeup, clothing, physical environments, voice, and performance, and then negotiated, broke, resisted, or refused them in relation to companies, audiences, platforms, and personal boundaries. We discuss implications for theory and platform design.



\bibliographystyle{ACM-Reference-Format}
\bibliography{citation}

\end{document}